\documentclass[reprint,amsmath,amssymb,aps]{revtex4-1}

\usepackage{CJKutf8}
\usepackage{graphicx}
\usepackage{xcolor}
\usepackage{hyperref}

\begin{document}

\title{Materials databases: the need for open, interoperable databases with standardized data and rich metadata}

\author{Fran\c{c}ois-Xavier Coudert}
 \email{fx.coudert@chimieparistech.psl.eu}
\affiliation{%
 Chimie ParisTech, PSL University, CNRS,
 Institut de Recherche de Chimie Paris, 75005 Paris, France
}

\date{\today}

\begin{abstract}
Driven by the recent rapid increase in the number of materials databases published (open and commercial), I discuss here some perspectives on the growing need for standardized, interoperable, open databases. The field of computational materials discovery is quickly expanding, and recent advances in data mining, high throughput screening, and machine learning highlight the potential of open databases.
\end{abstract}

\maketitle

One of the recent important trends in materials science is the emergence of several large-scale online databases of materials, trying to bring together experimental data with computational techniques in order to understand the behavior of materials families and design novel materials through data-mining? Maybe the most visible effort in this area is the Materials Project,\cite{Jain2013} a database of computed information on known and predicted properties, part of the US-funded Materials Genome Initiative\cite{White2012} launched in 2011. However, this trend is more general, and an increasing amount of research in the field focuses on the generation of these databases, their extension with additional data, and the use of these databases for analysis, screening, and prediction. This was clearly exemplified at recent materials meetings and materials modeling workshop, such as the MOFSIM 2019\cite{MOFSIM2019} meeting whose excellent discussion provoked this short comment.

\section{Existing databases}

The need for aggregation of curated data in the physical and chemical sciences was recognized very early, and possibly the best-known database in our field is the CRC Handbook of Chemistry and Physics\cite{CRC_handbook} (also known as the ``rubber book''), which has been published since 1914. When it comes to materials, databases ranging in size and scope have emerged since the beginning of computing, and have advanced at the same pace as both computer hardware and networking capabilities. The Cambridge Structural Database (CSD),\cite{CSD} launched in 1965, is one of the first numerical scientific databases, operating as a repository for validated experimental crystal structures of organic and organometallic compounds. Additional structural databases have emerged focusing on other categories of materials, including the Inorganic Crystal Structure Database (ICSD),\cite{Hellenbrandt2014} the Protein Data Bank (PDB),\cite{PDB} the American Mineralogist Crystal Structure Database,\cite{AMCSD} and the Crystallography Open Database (COD).\cite{Grazulis2012}

In addition to these structural databases, databases of materials properties have also been compiled, either by standardization institutes, learned societies, or commercial entities. We can cite examples of the NIST databases for materials and fluids properties, the glass property database SciGlass, Pearson's Crystal Data, or more specific examples like the Polymer Gas Separation Membrane Database from the Membrane Society of Australasia. Moreover, databases of hypothetical structures --- predicted by theory or computations --- have also appeared over time. In the field of porous materials, for example, databases of computed zeolitic structures have been published more than 15 years ago.\cite{Li2003, Earl2006} More recently, this effort has intensified --- probably in response to the increase in capacity of computations, as well as the ease of hosting large datasets online. Databases of various scales, containing hypothetical (enumerated) metal--organic frameworks, have been published.\cite{Wilmer2012} Other groups have worked on refining experimental databases to include additional data (such as atomic charges), making them suitable for computational applications and screening.\cite{Chung2014, Nazarian2016} All these databases, however, are hosted independently as archive files, with heterogeneous file formats, on individual research groups' websites.

We note, however, that there have been some recent initiatives in order to integrate data from different sources into larger, coherent databases. This is particularly the case of computational data, whose volume increases with high-performance computing capabilities. The goals differ for the various initiatives, but in general, they aim at providing large-scale platforms for open science and data sharing, as well as improve discoverability and searchability of existing data. A first example is the Materials Project,\cite{Jain2013} the aim of which is to ``remove guesswork from materials design in a variety of applications'' by computing properties of all known materials (and many that are not yet synthesized, too) through electronic structure analyses --- a project funded as part of the bigger Materials Genome Initiative\cite{White2012}. It aggregates structural data from other existing databases, as well as physical and chemical properties (band structures, elastic constants, piezoelectric tensors, electrode properties) computed as part of the project itself. Other platforms have been built to sharing results and resources in computational materials sciences, such as the ioChem-BD digital repository\cite{ioChemBD} and the more recent Materials Cloud\cite{MaterialsCloud}.

\section{The current state of affairs}

As shown above, the number of existing databases is increasing at a fast pace. Yet, the datasets themselves are often hosted using makeshift solutions, in different places. They are typically too large to be hosted as supporting information by the publisher of the associated research paper, and they might be expanded and refined over time, with the publication of updates not related to a particular peer-reviewed paper. In the most common case, they are hosted on the web server of the research group, on institutional repositories offered by some universities, or on free data hosting solutions such as Figshare,\cite{figshare} GitHub,\cite{github} etc. This leads to the available data being dispersed between several platforms, and in some cases, raises the question of long-term availability of the data (e.g., when it is hosted on a group web server or a commercial data hosting solution). Moreover, there exists no universal way to access these data, unlike what exists for open archives where protocols such as OAI-PMH have been developed for interoperability and discoverability of data sources.

This fragmentation of the landscape of materials databases is accompanied by a large heterogeneity in the formats used: while for crystalline material structures, the CIF (Crystallographic Information File) format is predominant, the data made available as part of the CIF file is not homogeneous between different groups. The use of symmetry operators, for example, is not always consistent, with some databases being stored with symmetry systematically lowered to $P1$. In addition to this heterogeneity in data format, there is also a general lack of availability of metadata, meaning that most of the databases do not contain information about how the data was generated, gathered, curated, and possibly updated. Yet, this metadata can be crucial in exploiting databases, in order to identify identical or related data items, to understand how databases evolve over time, and to allow further investigation of specific data items. Metadata enables researchers to answer simple queries such as: Where does this structure come from? Where was it first reported, and under what conditions was it synthesized? How was this computational property calculated? What are the conditions of reuse of this data?

\begin{figure}[t]
\includegraphics[width=0.8\linewidth]{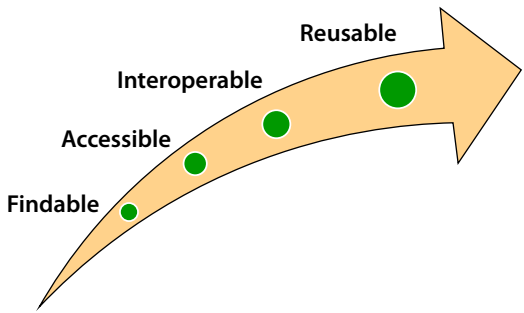}
\caption{FAIR data principles:\cite{Wilkinson2016} make data findable, accessible, interoperable and reusable.}
\label{fig:FAIR}
\end{figure}

Here, before highlighting some requirements for truly open materials databases, we want to introduce the FAIR data principles. The FAIR principles are a set of guidelines in order to make data findable, accessible, interoperable and reusable (Figure~\ref{fig:FAIR}). They have been formalized in 2016 by a consortium of scientists and organizations,\cite{Wilkinson2016} and were formally endorsed by the G20 Leaders at their 2016 summit in Hangzhou (\begin{CJK}{UTF8}{min}杭州\end{CJK}), China, in order to ``promote open science and facilitate appropriate access to publicly funded research results''.\cite{G20_Hangzhou} Requirements for FAIR data are:
\begin{itemize}
\item \textbf{Findable}: data are indexed in a searchable resource, has a stable unique identifier, and is described with rich metadata.
\item \textbf{Accessible}: data are retrievable using a standardized communications protocol, that is open, free, and universally implementable.
\item \textbf{Interoperable}: data are represented using an open, well-defined format; data and metadata are interlinked.
\item \textbf{Reusable}: data contains relevant metadata about its origin, a clear and accessible data usage license, and meet the community standards of its domain
\end{itemize}

\section{Requirements for open databases}

Based on these formal requirements, and during the discussions at MOFSIM 2019 and other workshops, there appears to be a need for more open and interoperable materials databases. We outline below some of the requirements, drawing both on the FAIR principles and shared experiences and discussions.

\textbf{Open databases.} A majority of the available data on materials is produced as part of academic research, and largely funded by public monies. Moreover, for data associated with public research, there is a general consensus that it should be accessible to all. Recent years have seen the addition of ``data availability'' requirements in several journals, and it is now a required part for any funding application. This creates a need for open databases, where content can be hosted regardless of its origin (contrary to institution-wide repositories) and where it is accessible to all. Moreover, the database should provide clear information about its users' rights when it comes to reusing the data, mining it, and republishing derived products.

Because open databases do not charge their users for subscription or access, they allow the dissemination of knowledge to categories of users that would otherwise find it difficult to obtain access: researchers in developing countries, nongovernmental organizations, independent researchers, journalists, even enthusiastic citizens. Operating such open databases of course requires funding, and it is important to note that a number of national and supranational initiatives have been launched in that direction (as discussed in the introduction).

\textbf{Interoperable databases.} It is relatively clear that, given the vastly different needs of scientists working in different areas of materials science, there can be no ``one size fits all'' database, i.e., no single centralized database that fulfills the needs of every different community. So, how can a good balance be reached in developing specific topical databases while retaining some uniformity, in order to avoid a complete fragmentation of the field? It turns out, this problem is one that has been worked on for many years in a related area, that of document (or papers) archives. While there are many different open archives on the internet, they have been developed in a way that allows interoperability between them. Specifically, the Open Archives Initiative (OAI) has standardized a Protocol for Metadata Harvesting (OAI-PMH) through which each archive exposes its metadata, in a common format, allowing for cross-database search and discoverability.\cite{OAIPMH}

In order to achieve this goal, several design choices are needed. One is the use of a well-documented, standardized Application Programming Interface (API). Through the use of that API, the data does not have to be retrieved with a database-specific client or web portal,\footnote{Although of course such clients can exist, providing a user-friendly way to query the database! The existence of a public API makes it possible for advanced users to develop their own portals, bringing added value to the database.} but can be written in any programming language without inside knowledge of how the database operates internally. This means, in turn, that code that is developed for one specific database will work seamlessly with all others.

Another is the inclusion of data in standard, publicly-documented file formats. Given that most current databases are currently structural databases, a part of this problem has already been addressed in the several past decades: crystal structures are uniformly reported in CIF format (although the details available are not always consistent), macromolecular structures are consistently in PDB format, etc. However, there is currently no unified format for storing the properties of these materials. This is made difficult by the fact that properties are rather diverse in their mathematical nature: some are dimensionless but others have units; some are integers or half-integers, others vary continuously; some are scalars, others are matrices or higher-order tensors. Moreover, they sometimes need to be accompanied by additional information: expected uncertainty, reference orientation, etc.

\textbf{With rich metadata and interlinked datasets.} Metadata can be defined, in its simplest form, as ``data that provides information about other data''. Like in any database, in a materials database metadata is crucial in assessing the data present, answering questions such as: How was this data gathered, by whom, when? In which conditions was a given property measured? For computational information, what was the theoretical method used, what is the level of description of the system? This is particularly important in databases of computational properties, where there can be a clear influence --- and sometimes even a systematic bias --- of the computational method chosen on the physical and chemical data calculated. If metadata is present in the databases, it opens the door to large-scale systematic explorations of various theoretical methods, and their comparison with experimental results obtained with different techniques, too.

Moreover, metadata can also provide much-needed links between several different interoperable databases. If a unique identifier is given for each dataset, and databases are interlinked through their metadata, it provides a much simpler exploration for users. It makes it easy to determine, e.g., if two properties from two datasets are independent or come from the same original calculation. It also allows greater discoverability, making it possible to find other properties in other databases, related to any given entry.

\textbf{Curation that preserves the scientific record.} The requirements listed above do not stop a fixed point of time, but instead must be considered throughout the databases' timeline. For example, metadata can record the time of measurement of a given data, but also its time of inclusion in the database, and its further history. Indeed, with any database of significant size, it is expected that curation of the data is an important topic, and the dataset will be modified to remove errors, updated to reflect new measurements, and sometimes data will be removed for a variety of legitimate reasons. However, for the sake of research reproducibility and conserving the scientific record, it is important that such modifications be recorded in the metadata --- just like corrections and retractions are publicly announced and archived for scientific articles. To my knowledge, this is not currently the case in existing databases, although the Materials Project is publishing ``release logs''\footnote{https://discuss.materialsproject.org/t/materials-project-database-release-log/1609} which are kept on a separate page, but not recorded in the database as metadata.

\textbf{Long-term availability.} Finally, this discussion cannot be concluded without addressing the issue of long-term availability of the deposited data, meaning that it is necessary, over time, to build institutional support with long-term commitments. This also requires planning for what happens if and when the hosting institutions decide to ``pull the plug'' on the project. There, having an open database with an API for direct access to bulk data is a benefit, because it means other interested parties can duplicate the database and take over hosting. Other lessons can be learned from open archives, and initiatives in that field such as CLOCKSS, a long-term preservation project for articles and books with highly-redundant mirroring.\cite{CLOCKSS}

\bibliography{references}

\end{document}